# Mean Number and Number Variance of Squeezed Coherent Photons In a Thermal State of Photons

Moorad Alexanian

*Department of Physics and Physical Oceanography*
*University of North Carolina Wilmington*
*Wilmington, NC 28403-5606*

**Abstract:** We consider the equilibrium thermal state of photons and determine the mean number and number variance of squeezed coherent photons. We use an integral representation for electro-magnetic radiation applicable both to systems in equilibrium and to systems in nonequilibrium to determine the spectral function of the radiation. The system considered is in thermal equilibrium and we find that the squeezed coherent photons are at a higher temperature than the photons themselves. Also, as expected, the mean number of squeezed coherent photons is greater than that of photons.

**Keywords:** thermal state, mean number, number variance, squeezed coherent photons, integral representation

## 1. Introduction

There are three basic states in quantum optics and in a recent paper [1], we dealt with the quantum Rabi oscillations via one- and two-photon transitions in two of these basic states, viz., coherent and squeezed coherent states. In this paper, we deal with the third state, viz., thermal states, and study the mean number and number variance of squeezed coherent photons in a thermal state of photons. In particular, one would expect the mean number of squeezed coherent states in a thermal state of photons to consist of the sum of three separate contributions, that is, the contribution from the thermal state, that from the coherent state, and that from the squeezed state. Indeed one does obtain such a sum of three terms; however, the contribution from the thermal state occurs with a temperature, which is higher than the temperature of the thermal state of photons. Squeezed coherent thermal states are studied with the aid of the Glauber second-order correlation functions [2]. Also, the Glauber *P*-representation for the squeezed thermal state are calculated and compared to the *P*-representation for the squeezed states [2]. Thermal coherent states have also been used to generalize the usual (zero-temperature) *P*- and *Q*- representations of operators to their nonzero temperature counterparts [3]. This paper is arranged as follows. In Sec. 2, we review the use of the Bogoliubov-Valatin transformation to generate the creation and annihilation operators for the squeezed coherent photons. In Sec. 3 A, we calculate the mean number and number variance of photons in a thermal state of squeezed coherent photons. In Sec. 3 B, we calculate the mean number and number variance of squeezed coherent photons in a thermal state of photons. In Sec. 4, we review the integral representation for the mean number of bosons in thermal equilibrium and evaluate the spectral function to the cases considered in Sec. 3. Finally, Sec. 5 summarizes our results.

## 2. Bogoliubov-Valatin Transformation

In a recent paper [1], we considered the Bogoliubov-Valatin [4, 5] canonical transformation, which is in general not unitary, to generate quasiparticles

$$\widehat{A} = \beta \hat{a} + \gamma \hat{a}^\dagger \qquad \widehat{A}^\dagger = \beta^* \hat{a}^\dagger + \gamma^* \hat{a} \tag{1}$$

where $\hat{a}$, $\hat{a}^\dagger$ are the photon annihilation and creation operators, respectively, with vacuum state $\hat{a}|0\rangle = 0$. The creation and annihilation operators $\widehat{A}$ and $\widehat{A}^\dagger$ satisfy the communication relation

$$[\widehat{A}, \widehat{A}^\dagger] = 1 \tag{2}$$

provided $|\beta|^2 - |\gamma|^2 = 1$.

The corresponding normalized vacuum state, viz., $\widehat{A}|0\rangle = 0$, is given by [1]

$$|0\rangle = \hat{S}(\zeta)|0\rangle \tag{3}$$

where

$$\hat{S}(\zeta) = exp\left(-\frac{\zeta}{2}\hat{a}^{\dagger 2} + \frac{\zeta^*}{2}\hat{a}^2\right) \tag{4}$$

is the squeezing operator with

$$\frac{\gamma}{\beta} = e^{i\varphi} \tanh(r) \tag{5}$$

and $\zeta = r exp(i\varphi)$.

The non-unitary transformation (1) contains three real variables. One can reduce the number of variables to only two real variables and thus make the transformation (1) unitary by choosing

$$\beta = \cosh(r) \qquad \gamma = e^{i\varphi} \sinh(r) \tag{6}$$

and so one obtains from (1) the unitary transformation

$$\hat{S}(\zeta)\hat{a}\hat{S}(-\zeta) = \cosh(r)\,\hat{a} + e^{i\varphi}\sinh(r)\,\hat{a}^\dagger = \widehat{A} \tag{7}$$

$$\hat{S}(\zeta)\hat{a}^\dagger\hat{S}(-\zeta) = \cosh(r)\,\hat{a}^\dagger + e^{-i\varphi}\sinh(r)\,\hat{a} = \hat{A}^\dagger \qquad (8)$$

One can generate a coherent state for the quasiparticles [1] by the action on the vacuum $|0\rangle$ of the Glauber displacement operator

$$\hat{D}(\alpha)|0\rangle = exp(\alpha\hat{A}^\dagger - \alpha^*\hat{A})|0\rangle = \hat{S}(\zeta)\hat{D}(\alpha)|0\rangle, \qquad (9)$$

where $\alpha = |\alpha|\exp(i\theta)$. Therefore, the coherent state of quasiparticles is the squeezed coherent state of photons.

The creation and annihilation operators $\hat{B}$ and $\hat{B}^\dagger$, respectively, for the squeezed coherent photons are given by

$$\hat{B} = \hat{S}(\zeta)\hat{D}(\alpha)\hat{a}\hat{D}(-\alpha)\hat{S}(-\zeta) = \cosh(r)\,\hat{a} + e^{i\varphi}\sinh(r)\,\hat{a}^\dagger - \alpha \qquad (10)$$

and

$$\hat{B}^\dagger = \hat{S}(\zeta)\hat{D}(\alpha)\hat{a}^\dagger\hat{D}(-\alpha)\hat{S}(-\zeta) = e^{-i\varphi}\sinh(r)\,\hat{a} + \cosh(r)\,\hat{a}^\dagger - \alpha^*, \qquad (11)$$

with inverses

$$\hat{a} = \hat{D}(-\alpha)\hat{S}(-\zeta)\hat{B}\hat{S}(\zeta)\hat{D}(\alpha) = \cosh(r)\,\hat{B} - e^{i\varphi}\sinh(r)\,\hat{B}^\dagger + \alpha\cosh(r) - \alpha^*e^{i\varphi}\sinh(r) \quad (12)$$

and

$$\hat{a}^\dagger = \hat{D}(-\alpha)\hat{S}(-\zeta)\hat{B}^\dagger\hat{S}(\zeta)\hat{D}(\alpha) = -e^{-i\varphi}\sinh(r)\,\hat{B} + \cosh(r)\,\hat{B}^\dagger + \alpha^*\cosh(r) - \alpha e^{-i\varphi}\sinh(r). \qquad (13)$$

### 3. Mean Number and Number Variance

We evaluate the mean number and number variance of photons in the thermal state of squeezed coherent photons and the mean number and number variance of squeezed coherent photons in the thermal state of photons. Actually, the two results are related via a transformation of parameters as shown in Sec. 3 B.

#### A. Thermal state of squeezed coherent photons

We consider the thermal equilibrium state of an ideal gas of squeezed coherent photons with density matrix

$$\hat{\rho}_B = \frac{\exp(-\beta_B\hbar\omega\hat{B}^\dagger\hat{B})}{Tr[\exp(-\beta_B\hbar\omega\hat{B}^\dagger\hat{B})]}, \qquad (14)$$

where $\beta_B = (k_B T)^{-1}$, $k_B$ being Boltzmann's constant and $T$ being the absolute temperature. The thermal average of operator $\hat{O}$ is denoted by

$$\langle \hat{\rho}_B \hat{O} \rangle \equiv \langle \hat{O} \rangle_B. \tag{15}$$

We obtain for the mean photon number in the thermal state of squeezed coherent photons

$$\langle \hat{a}^\dagger \hat{a} \rangle_B = \frac{\cosh(2r)}{e^{\beta_B \hbar \omega} - 1} + \sinh^2(r) + |\alpha \cosh(r) - \alpha^* \sinh(r) e^{i\varphi}|^2, \tag{16}$$

with the aid of (12) and (13).

The photon number variance follows from (12) and (13) and gives

$$\Delta n^2 = \langle (\hat{a}^\dagger \hat{a})^2 \rangle_B - (\langle \hat{a}^\dagger \hat{a} \rangle_B)^2 = \frac{\cosh(4r)}{(e^{\beta_B \hbar \omega} - 1)^2} + \frac{\cosh(4r) + 2|\alpha \cosh(2r) - \alpha^* \sinh(2r) e^{i\varphi}|^2}{(e^{\beta_B \hbar \omega} - 1)}$$
$$+ \frac{1}{2} \sinh^2(2r) + |\alpha \cosh(2r) - \alpha^* \sinh(2r) e^{i\varphi}|^2 \tag{17}$$

Both results for the mean photon number and variance lead, in the limit $T \to 0$, to the results given by Eqs. (18) and (19), respectively, in Ref. (1) for $\varphi = 2\theta$. Also, one obtains the results for the ideal photon gas when $\alpha = 0$ and $r = 0$, viz., $\Delta n^2 = \bar{n}(\bar{n} + 1)$, where $\bar{n} = [e^{\beta_B \hbar \omega} - 1]^{-1}$.

### B. Thermal state of photons

We consider the thermal equilibrium state of an ideal gas of photons with density matrix

$$\hat{\rho}_a = \frac{\exp(-\beta_B \hbar \omega \hat{a}^\dagger \hat{a})}{Tr[\exp(-\beta_B \hbar \omega \hat{a}^\dagger \hat{a})]}, \tag{18}$$

One can easily evaluate the mean number of squeezed coherent photons in the thermal state of photons, viz.,

$$\langle \hat{B}^\dagger \hat{B} \rangle_a = \frac{Tr[\hat{B}^\dagger \hat{B} e^{-\beta_B \hbar \omega \hat{a}^\dagger \hat{a}}]}{Tr[e^{-\beta_B \hbar \omega \hat{a}^\dagger \hat{a}}]} = \frac{\cosh(2r)}{e^{\beta_B \hbar \omega} - 1} + \sinh^2(r) + |\alpha|^2 \tag{19}$$

with aid of (10) and (11).

The number variance of squeezed coherent photons follows from (10) and (11) yielding

$$\Delta n^2 = \langle (\hat{B}^\dagger \hat{B})^2 \rangle_a - \left( \langle \hat{B}^\dagger \hat{B} \rangle_a \right)^2 = \frac{\cosh(4r)}{(e^{\beta_B \hbar \omega} - 1)^2} + \frac{\cosh(4r) + 2|\alpha \cosh(r) + \alpha^* \sinh(r) e^{i\varphi}|^2}{(e^{\beta_B \hbar \omega} - 1)}$$
$$+ \frac{1}{2} \sinh^2(2r) + |\alpha \cosh(r) + \alpha^* \sinh(r) e^{i\varphi}|^2. \tag{20}$$

The squeezed coherent photon of Sec. 3 A is given by $\hat{S}(\zeta)\hat{D}(\alpha)|0\rangle$. The results in this Sec. 3 B can be equally obtained as in Sec. 3 A but with a "squeezed coherent photon" given by the adjoint, that is, $\left(\hat{S}(\zeta)\hat{D}(\alpha)\right)^\dagger |0\rangle = \hat{D}(-\alpha)\hat{S}(-\zeta)|0\rangle$ instead. In fact, we obtain the results (16) and (17) from the results (19) and (20) under the transformation

$$\alpha \to \alpha \cosh(r) - \alpha^* \sinh(r) e^{i\varphi} \qquad \text{and} \qquad \zeta \to -\zeta. \tag{21}$$

## 4. Integral Representation

Sometime ago, an integral representation for an interacting or a non-equilibrium Bose gas was derived [6, 7]. The integral representation of a Bose gas in thermal equilibrium at temperature $T$ is given by

$$n(\omega, T) = \int_0^\infty dT' \int_0^\infty d\mu \frac{\sigma(T, T', \mu)}{e^\mu e^{\hbar \omega / k_B T'} - 1}, \tag{22}$$

where $\mu$ is the chemical potential and the spectral function $\sigma(T, T', \mu) \geq 0$ and can be no more singular than a Dirac $\delta$-function. In the nonequilibrium case, the time variable $t$ must be included in the arguments of both functions, viz., $n(\omega, T, t)$ and $\sigma(T, T', \mu, t)$. If the contribution to the radiation field (22) arises from a single source, then one has the normalization condition

$$\int_0^\infty dT' \int_0^\infty d\mu \, \sigma(T, T', \mu) = 1. \tag{23}$$

However, in the case of multiple sources, the integral (23) is greater than unity. This occurs, for instance, in the description of the cosmic background radiation that deviates from a strictly blackbody that requires an additional source other than the remnant of the big bang [8].

If the number of bosons, for instance, in the case of photons, is not conserved then

$$n(\omega, T) = \int_0^\infty dT' \frac{\sigma(T, T')}{e^{\hbar \omega / k_B T'} - 1}. \tag{24}$$

Note that the low frequency photons in (24) are in thermal equilibrium, owing to Bremsstrahlung, with temperature

$$T_{eq}(T) = \int_0^\infty dT' \ T'\sigma(T,T'). \tag{25}$$

The first term in both expressions for the mean photon number in (16) and (19) are described by (22). One has for both (16) and (19) that

$$\sigma(T,T',\mu) = \cosh(2r)\,\delta(T'-T)\delta(\mu). \tag{26}$$

Therefore, the low-frequency squeezed coherent photons are in thermal equilibrium with $T_{eq} = \cosh(2r)T \geq T$, that is, a higher temperature than the photon temperature $T$. However, the remaining terms in both (16) and (19) are described by (22) with a non-zero chemical potential albeit infinite temperature.

Therefore, one has for the spectral function for (16)

$$\sigma_B(T,T',\mu) = \cosh(2r)\,\delta(T'-T)\delta(\mu) + \delta(\mu - \mu_B)\delta(T' - \infty), \tag{27}$$

where

$$\mu_B = \ln\left(\frac{\cosh^2(r) + |\alpha\cosh(r) - \alpha^*\sinh(r)e^{i\varphi}|^2}{\sinh^2(r) + |\alpha\cosh(r) - \alpha^*\sinh(r)e^{i\varphi}|^2}\right). \tag{28}$$

Similarly, one has for the spectral function for (19)

$$\sigma_a(T,T',\mu) = \cosh(2r)\,\delta(T'-T)\delta(\mu) + \delta(\mu - \mu_a)\delta(T' - \infty), \tag{29}$$

where

$$\mu_a = \ln\left(\frac{\cosh^2(r) + |\alpha|^2}{\sinh^2(r) + |\alpha|^2}\right). \tag{30}$$

It is curious that the low frequency, squeezed photons are at a higher temperature $T\cosh(2r)$ than the photons in their own thermal equilibrium state with temperature $T$. Therefore, the squeezed coherent photons are not only more numerous than the photons but the low-frequency squeezed coherent photons are at a higher temperature than the photon temperature. The effect of coherent photons, as expected, is only to increase the number of photons.

## 5. Summary and Discussion

The calculation of the mean number of squeezed coherent photons in a thermal state of photons is analyzed vis-á-vis an integral representation for the mean number of bosons. The contributions from the integral give rise to two terms one where the number of bosons are not conserved, zero chemical potential, and a second term where the number of bosons are conserved, nonzero chemical potential. The latter resembles somewhat the condensation of photons in the zero frequency state. It is interesting that the temperature of the squeezed coherent photons is greater than the temperature of the photon gas.